\documentclass[conference]{IEEEtran}
\IEEEoverridecommandlockouts
\ifCLASSINFOpdf

\else

\fi

\usepackage[font=footnotesize]{caption}
\usepackage{graphicx}
\usepackage{subcaption}
\usepackage{amsthm}
\usepackage{mathrsfs}  
\usepackage{amsmath,amssymb,amsthm,amsbsy}
\usepackage{mathtools}
\usepackage{commath}
\usepackage{breqn}
\usepackage{algorithmic}
\usepackage{textcomp}
\usepackage{xcolor}
\usepackage{url}
\usepackage{float}
\usepackage{multirow}
\usepackage{makecell}
\usepackage{cuted}

\usepackage[style = ieee]{biblatex}
\bibliography{main}
\begin{document}

\title{Extending the Dissipating Energy Flow Method to Flexible AC Transmission Systems
}

\author{Kaustav~Chatterjee, Sayan Samanta, and Nilanjan Ray Chaudhuri\\
\IEEEauthorblockA{School of
of Electrical Engineering and Computer Science \\
The Pennsylvania State University\\
kuc760@psu.edu, ~sps6260@psu.edu, ~nuc88@psu.edu}
\vspace{-0.3cm}
}

\maketitle

\begin{abstract}
In recent years, the dissipating energy flow (DEF) method has emerged as a promising tool for online localization of oscillation sources. In literature, the mathematical foundations of this method is well-studied for networks with synchronous generators. In this paper, we extend the analysis to felxible AC transmission systems (FACTS). To that end, we derive the DEF-expressions for a thyristor controlled series capacitor (TCSC) and a static synchronous compensator (STATCOM) operating with conventional control strategies. Analyzing their respective DEFs, we obtain the conditions for which these FACTS devices could behave as the sources of oscillation energy. Our findings are structured into propositions and are supported through numerical case studies on IEEE test systems.  
\end{abstract}


\section{Introduction}

Sustained low-frequency oscillations may appear in a power system due to the poorly-damped natural modes or as a result of an external periodic disturbance. Such oscillations induce mechanical stress on the turbines, limit utilization of existing transmission assets causing congestion, and in the worst case, if not adequately damped, may lead to large-scale outages. An important step in mitigating these is to identify the sources either responsible for forcing these oscillations or contributing towards the deterioration of the system damping. DEF-based oscillation source localization, as proposed in \cite{tsingua1}, is one such method which has shown significant promise when tested on real-world oscillation problems \cite{masala}. Derived from the theory of transient energy functions, the method characterizes a power system component as a source or a sink of oscillation energy depending on the slope of its DEF integral \cite{tsingua1}. 
Despite its reasonable success, it is important to note that, the DEF method may perform poorly when the underlying modeling assumptions (see, \cite{tsingua1}) are violated \cite{maniDEF}. 

\par Moreover, while for forced oscillations, it is relatively easier to singularly identify an element as a source from the high-value positive slope of its DEF,\footnote{usually the amount of energy injected by the forcing function is much higher than any other element} it is more challenging for a poorly-damped natural oscillation. In most cases, there can be more than one synchronous generator which, depending on their damping torque characteristics and observability of the mode, may appear as potential sources of the oscillation. This can be further complicated by positive-sloped DEFs that may originate from other controllable network elements like $-$ FACTS, high voltage direct current (HVDC) systems, and converter-interfaced renewable energy sources. 
While the theoretical connections between the DEF and damping contribution from synchronous generators is well-studied in literature \cite{chen2, sam_tpwrs, LCSS}, no insights have been developed relating the role of the control schemes in power electronics-interfaced devices in shaping their respective DEFs. 

To that end, in this paper, we extend the mathematical analysis of the DEF to study the oscillation-sink/source behavior of the FACTS devices and their controls. We derive the transient energy functions for 1) a TCSC, and 2) a STATCOM. For each device, we then compare among the possible control schemes to check for the presence (or absence) of path-dependent integrals in their respective energy functions. Since the existence of a path-dependent terms implies transient energy dissipation (or production), we can then build on this to mathematically explain the nature of their DEFs.  Establishing this link between the control law, the path-dependent transient energy integral, and the slope of the DEF serves two purposes: first, it helps us appreciate how a control law influences the source/sink behavior of a given device, and second, it gives us an intuitive understanding (if possible) of how to modify the control law and tune the controller parameters to change the nature of the device from source to sink of oscillation energy. 

\par The paper is organized as follows: in Section \ref{tcsc_sec}, we derive and analyze the DEF-expressions for a TCSC operating under 1) a  fixed compensation, and 2) variable compensation using a damping controller in closed-loop feedback. In Section \ref{stat_sec}, we study the DEF-expression for a STATCOM operating under 1) constant current control, and 2) reactive power $-$ voltage droop control. For each of these, the numerical studies on a modified IEEE 2-area 4-machine system are presented in their respective sections. Finally, the conclusions are discussed in Section \ref{conclusions}.

\section{DEF Analysis of a TCSC}
\label{tcsc_sec}
Assume a power network, wherein between nodes $i$ and $k$, the branch $L_{ik}$ is a TCSC. The admittance of the branch be, 
\begin{equation}\small
\vspace{-0.1cm}
   y_{ik} = jb = j(1 - k_c)b_0
\end{equation}
where, $b$ is the variable reactance of the TCSC and $k_c$ is the state variable modulating this reactance over its nominal value $b_0$. Further, $k_c = k_{c_0} + \Delta k_c$, where, $k_{c_0}$ is the value in the steady-state and $\Delta k_c$ is the perturbation under transients.  
\par As derived in \cite{tsingua1}, the transient energy flowing into the branch $L_{ik}$ from bus $i$ is $\int \Im( \vec{I}^*_{ik} d\vec{V_i})$, where $\vec{I}_{ik}$ and $\vec{V_i}$ are respectively the complex-valued line current and bus voltage phasors in the network reference frame, and $\Im(\cdot)$ denotes the imaginary part of a complex number. Similarly, the energy flowing into branch $L_{ik}$ from bus $k$ is given by $\int \Im( \vec{I}^*_{ki} d\vec{V_k})$. Combining these two, the total energy flowing into the TCSC, denoted by $W_{\text{TCSC}}$, can be written as,
\begin{equation}\small
\label{Wik1}
  W_{\text{TCSC}} =  \int \Im( \vec{I}^*_{ik} d\vec{V_i} + \vec{I}^*_{ki} d\vec{V_k} ) = \int \Im\Big( \vec{I}^*_{ik} (d\vec{V_i} - d\vec{V_k}) \Big).
\end{equation}
Denoting $\vec{V}_{ik} = \vec{V}_i - \vec{V}_k$, and thereafter, substituting $\vec{I}_{ik} = y_{ik}\vec{V}_{ik}$ into (\ref{Wik1}), we get
\begin{equation}\small
    \label{Wik2}
    W_{\text{TCSC}} =  \int \Im\Big( y^*_{ik}~\vec{V}^*_{ik}~ d\vec{V}_{ik} \Big)  = b_0\int \Im\Big(-j (1 - k_c)~\vec{V}^*_{ik}~ d\vec{V}_{ik} \Big).
\end{equation}
Next, let us define, $\vec{V}_{ik} = U \text{e}^{j \phi} = U_x + j U_y$ wherein, $U = U_0 + \Delta U$. Following which, (\ref{Wik2}) reduces to
\begin{equation}\label{Wik3} \small
\begin{aligned}
    W_{\text{TCSC}} &= b_0 \int \Im\Big( -j(1 - k_c)~U \text{e}^{-j \phi} ~(dU + j U d\phi) \text{e}^{j \phi} \Big) \\
    &= -~ b_0 \int (1-k_c)~U~dU \\
    &= \bigg(-b_0  (1-k_{c_0})~U_0\int d\Delta U\bigg) \\&+ \bigg(-b_0~(1-k_{c_0})\int \Delta U~d\Delta U
    + b_0 \int \Delta k_c ~U ~d\Delta U \bigg)
    \\&\overset{\Delta}{=} W_{\text{TCSC}}^0 + W_{\text{TCSC}}^D .
\end{aligned}
\end{equation}
If the evaluation of $W_{\text{TCSC}}$ depends on the path traversed by the variable $\Delta U$, instead of the initial and final values of $\Delta U$ alone, we call the integral {path-dependent}. Else, it is termed as path-independent. Path-dependence implies, the expression of $W_{\text{TCSC}}$ contains the monotonically changing component related to energy dissipation or production in the device. Observe that, 
\begin{equation*}\label{wtcsc_def}\small
\begin{aligned}
    W_{\text{TCSC}}^0(t_1) = - b_0 (1-k_{c_0})U_0\int_0^{^{\Delta U(t_1)}} \hspace{-0.8cm}d\Delta U = - b_0  (1-k_{c_0})U_0 ~\Delta U(t_1)
    \end{aligned}
\end{equation*}
is path-independent. Also, in
\begin{equation*}\label{wtcsc_def}\small
\begin{aligned}
    &W_{\text{TCSC}}^D(t_1) = - b_0~(1-k_{c_0}) \int_0^{^{\Delta U(t_1)}} \hspace{-0.8cm}\Delta U~d\Delta U + b_0 \int_0^{^{\Delta U(t_1)}} \hspace{-0.8cm}\Delta k_c ~U ~d\Delta U\\
    &= -\frac{1}{2}~b_0~(1-k_{c_0})\{\Delta U(t_1)\}^2 ~+~ b_0 \int_0^{^{\Delta U(t_1)}}\hspace{-0.7cm} \Delta k_c ~U ~d\Delta U,
    \end{aligned}
\end{equation*}
the first term is always path-independent and represents the change in the stored energy of the TCSC. 
Further, if $\Delta k_c =0$ (implying, fixed compensation, $k_c = k_{c_0}$) or $\Delta k_c$ is an explicit algebraic function of $\Delta U$ (e.g., $\Delta k_c = K_p \Delta U$ 
  output of a proportional controller with $\Delta U$ as feedback), the second term in $W_{\text{TCSC}}^D$ is also path-independent. Absence of path-dependent terms imply there is no generation or dissipation of oscillation energy. This is summarized in the propositions below.  

\vspace{0.2cm}
\textbf{Proposition 1.} \emph{A TCSC-compensated lossless\footnote{~for a resistive element, the slope of its DEF is indefinite (see, \cite{chen_res})} transmission line with fixed compensation is neither a source nor a sink of oscillation energy. }

\textbf{Proposition 2.} \emph{A TCSC modulating the reactance of a  transmission line as an explicit algebraic function of the difference in its (i.e., TCSC's) terminal bus voltage magnitudes is neither a source nor a sink of oscillation energy. }

\par
\begin{figure}[h]
    \centering
    \vspace{-0.1cm}
    \includegraphics[width=\linewidth]{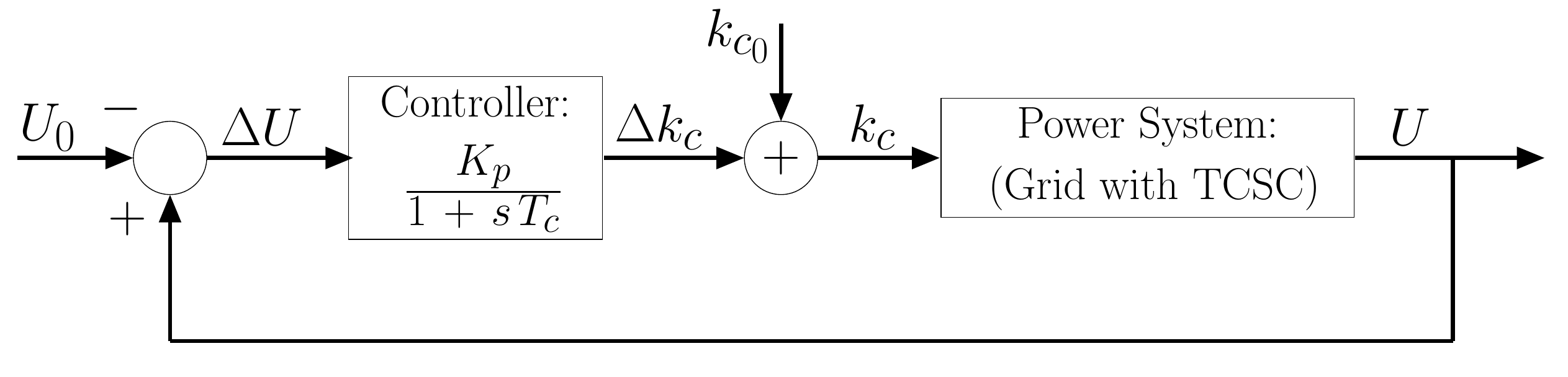}
    \caption{Closed-loop control for modulating $k_c$ of a TCSC.}
    \vspace{-0.1cm}
    \label{controller}
\end{figure}

 However, when $k_c$ is an implicit function involving the time-variable $t$ alongside $U(t)$, $W_{\text{TCSC}}^D(t_1)$ may become path dependent. To demonstrate this, consider the closed-loop system in Fig \ref{controller}.
Unlike before, where $\Delta k_c$ was an algebraic function of $\Delta U$, we now have a controller, with frequency-dependent gain and phase, modulating $\Delta k_c$ as shown below.
\begin{equation}\small
\label{kc}
    \Delta \dot{k}_c = -\frac{1}{T_c} \Delta k_c + \frac{K_p}{T_c} \Delta U.
    \vspace{-0.1cm}
\end{equation}
The solution to the state equation in (\ref{kc}) is expressed as
\begin{equation}\small
    \label{kct}
    \Delta k_c(t) = \text{e}^{-\frac{t}{T_c}} \Delta k_c(0) + \frac{K_p}{T_c} \int_{0}^{t} \text{e}^{-\frac{t-\tau}{T_c}} \Delta U(\tau) ~ d\tau .
    \vspace{-0.2cm}
\end{equation}
Since,  $\Delta k_c(0) = 0$, $\Delta k_c(t)$ only comprises of the forced response. Substituting this in the expression of $W_{\text{TCSC}}^D(t_1)$,
\begin{equation}\label{wtcsc_def_2} \small
\begin{aligned}
    &W_{\text{TCSC}}^D(t_1) = -\frac{1}{2}~b_0~ (1-k_{c_0}) \{\Delta U(t_1)\}^2 \\ &+ b_0~\frac{K_p}{T_c} \int_0 ^{t_1} \int_0 ^t  \text{e}^{-\frac{t-\tau}{T_c}} \Delta U(\tau) \Big(U_0 + \Delta U(t)\Big) ~ \Delta\dot{U}(t) ~d\tau ~dt .
    \end{aligned}
\end{equation}
Our claim is that, between any two time-instants, with same initial and final system states, the evaluation of the second term in $W_{\text{TCSC}}^D$ (see, (\ref{wtcsc_def_2})) depends on the trajectory traced by $\Delta U(t)$ (or $U(t)$). In other words, for different paths of $U(t)$ between the same $U_0$ and $U(t_1)$, the value of $W_{\text{TCSC}}^D(t_1)$ depends on the non-conservative contributions in the respective paths. This is summarized in the proposition below.
\vspace{0.1cm}
\par \textbf{Proposition 3.} \emph{A controller with frequency-dependent gain and phase
modulating the control input} $k_c$, \emph{renders} $W_{\text{TCSC}}^D$ \emph{path-dependent.}

\par
\begin{figure}[h]
\vspace{-0.4cm}
  \begin{subfigure}{.5\columnwidth}
    \centering
    \includegraphics[width=1.05\linewidth]{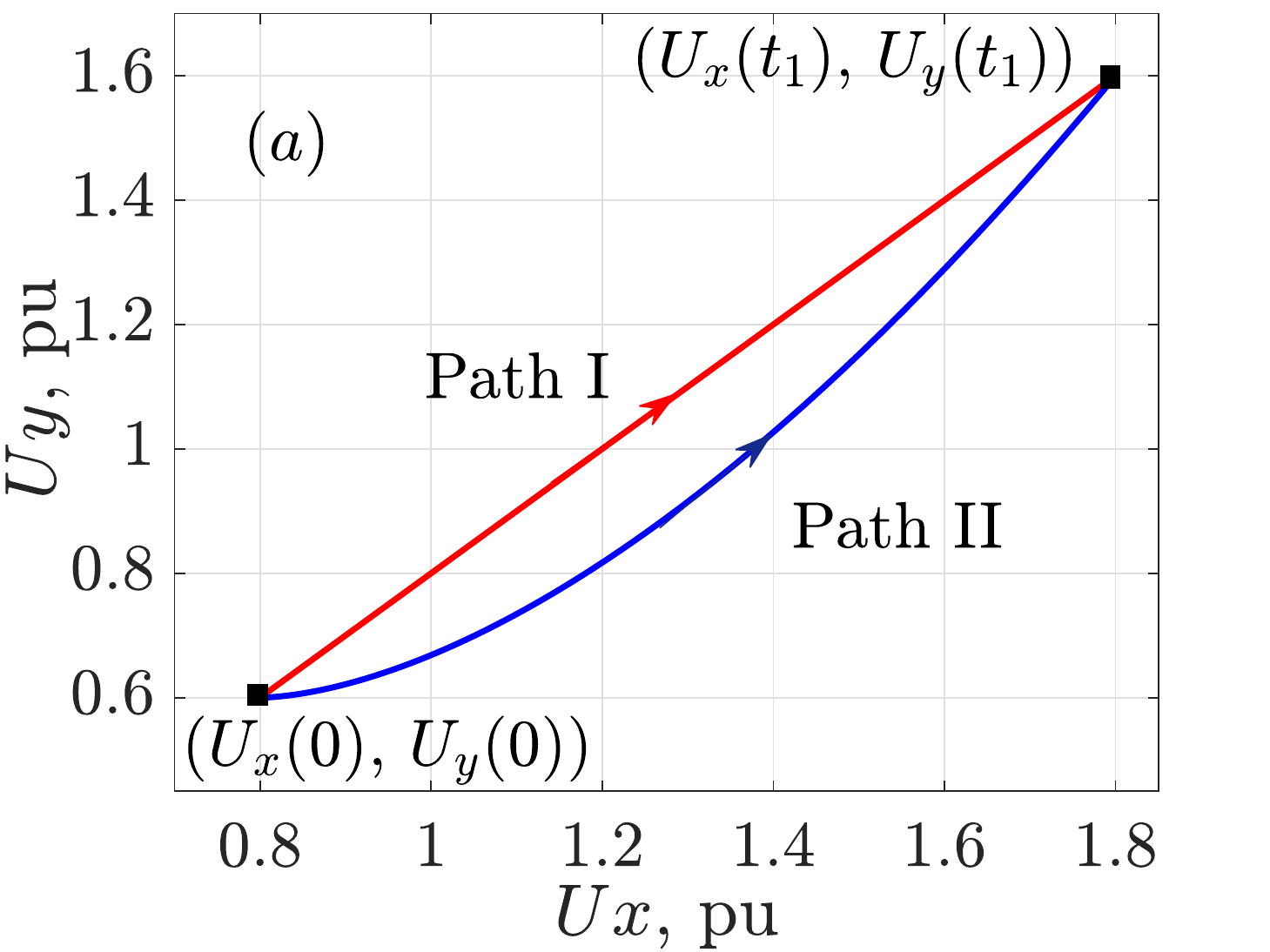}
  \end{subfigure}%
  \begin{subfigure}{.5\columnwidth}
    \centering
    \includegraphics[width=1.05\linewidth]{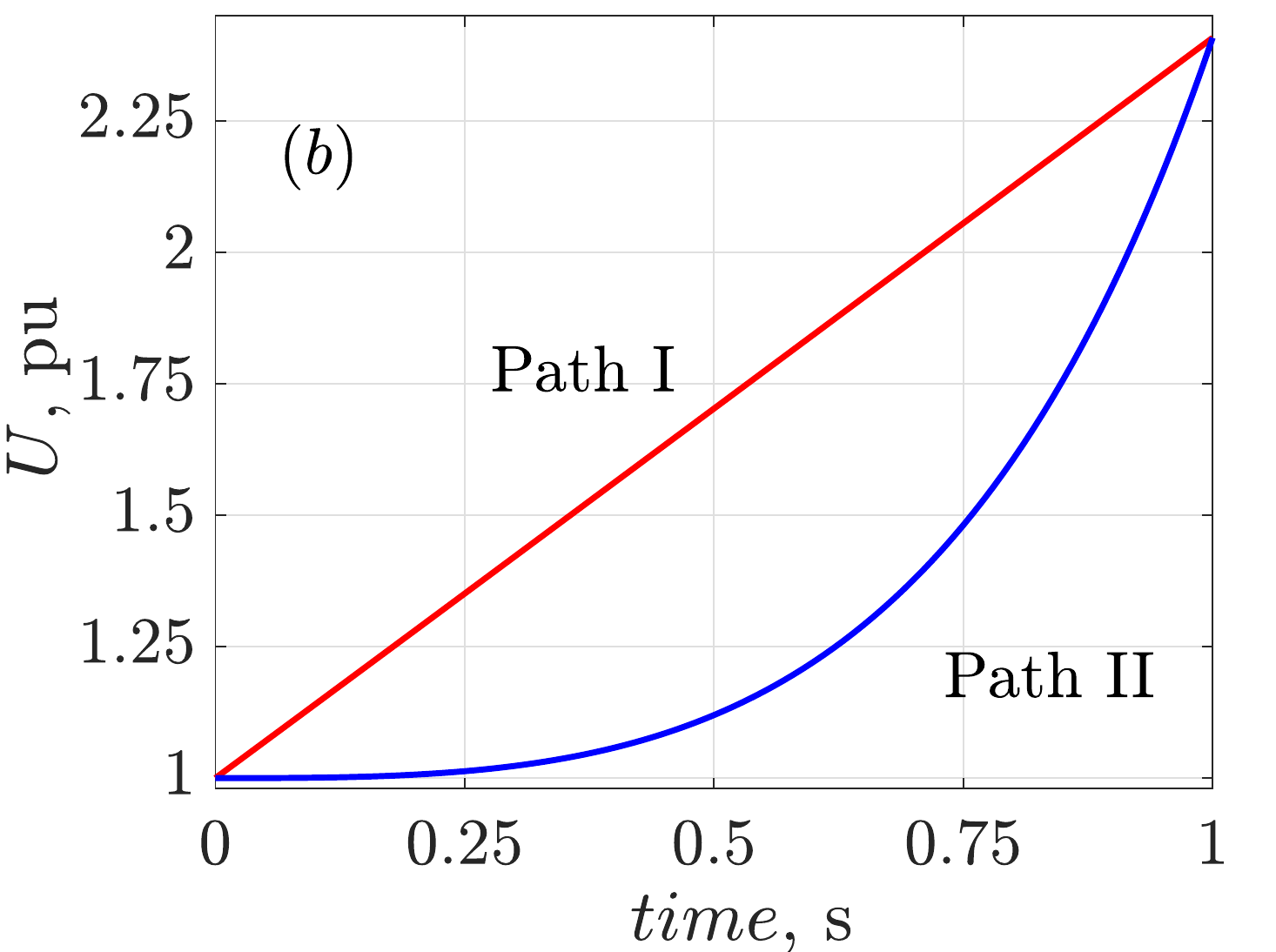}
  \end{subfigure}%
  \caption{Two different paths between initial and final values of $U(t)$ for $\alpha=1$ (i.e., final value at $t_1 = 1$ s) in (a) $U_x-U_y$ plane and (b) $t-U(t)$ plane. }
  \label{Upath}
  \vspace{-0.6cm}
  \end{figure}

 We verify this proposition using a simple numerical study. In (\ref{wtcsc_def_2}), assume, $U(0) = U_0 = 1$ pu with $U_x(0) = 0.8$ pu and $U_y(0) = 0.6$ pu. Next, consider two different paths between time instants $t = 0$ and $t = t_1$ s as shown below.
\begin{equation*}
   \begin{aligned}
       &\text{Path I:}~~ U_x(t) = U_x(0) + \alpha t~;~ U_y(t) = U_y(0) + \alpha t\\
       &\text{Path II:}~ U_x(t) = U_x(0) + \alpha^k t^k~;~ U_y(t) = U_y(0) + \alpha^n t^n
   \end{aligned} 
\end{equation*}
\noindent In the $U_x - U_y$ plane, path I traces the straight line $U_y = U_x + \big( U_y(0) - U_x(0)\big)$, while path II traces a nonlinear trajectory given by $U_y = \big(U_x - U_x(0)\big)^{\frac{n}{k}} + U_y(0)$. This is shown in Fig \ref{Upath}$(a)$. To ensure that the final values are same for both the paths, we choose $t_1 = \frac{1}{\alpha}$. Further, for this numerical study, we arbitrarily set $k =3$ and $n = 5$. The time-domain trajectories of $U(t)$ for paths I and II are shown in Fig \ref{Upath}$(b)$. 
\par
\begin{figure}[H]
    \hspace{0.2cm}
    \includegraphics[width=\linewidth]{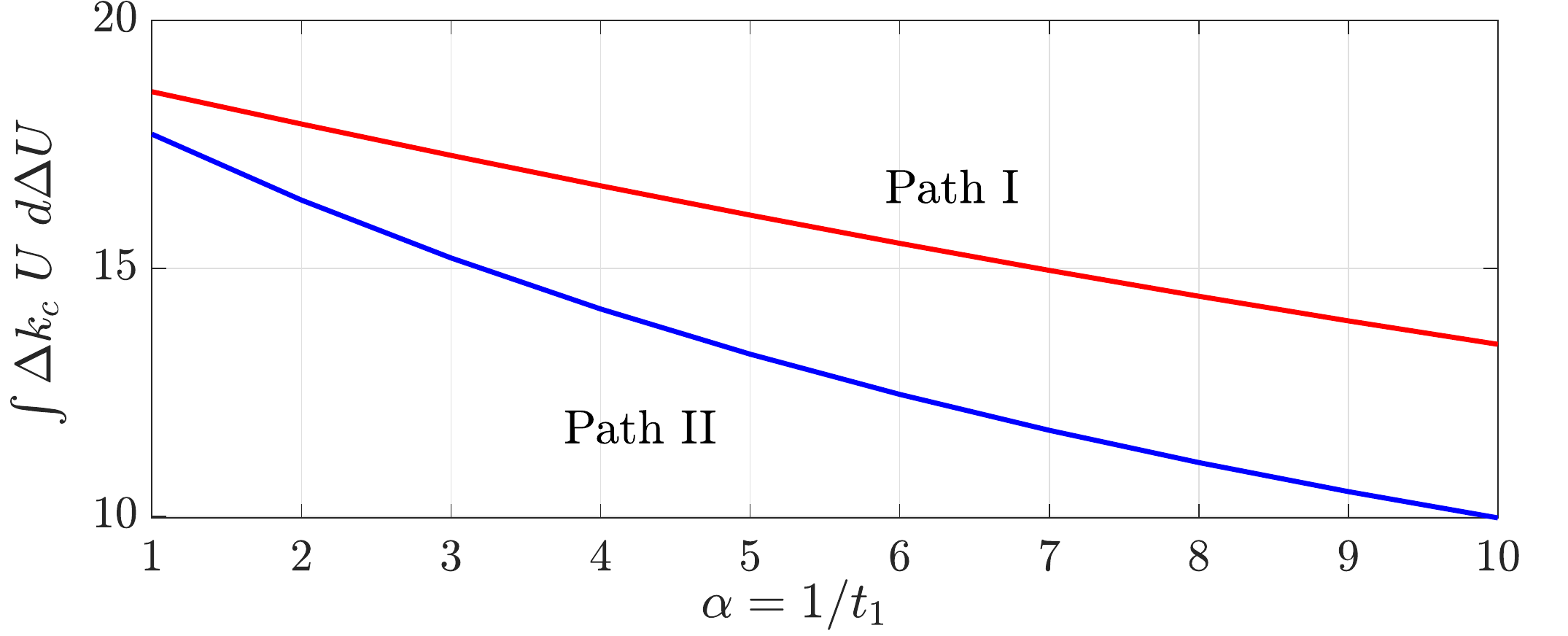}
    \caption{Path-dependence of the second integral in $W_{\text{TCSC}}^D$ for the two different paths in Fig  \ref{Upath} corresponding to different $\alpha$ values.}
    \label{Galpha}
    \vspace{-0.2cm}
\end{figure}

  
Since, the area enclosed by $U(t) \big(= U_0 + \Delta U(t)\big)$ in path I is different from that of path II (see, Fig  \ref{Upath}$(b)$), the evaluation of the second term in (\ref{wtcsc_def_2}) at any time $t_1$ is dependent on the path traced by $U(t)$ between $t= 0$ and $t_1$. This is shown in Fig  \ref{Galpha} $-$ the integral in the second term of (\ref{wtcsc_def_2}) evaluated for different values of $\alpha$ (and therefore, different endpoints $t = t_1$) for both paths I and II are different. Path-dependence of the second term implies, $W_{\text{TCSC}}^D$ as a whole is also path-dependent.
\par The importance of proposition 3, in the context of oscillation monitoring and source localization, is discussed next. The path dependence of the energy function $W_{\text{TCSC}}^D$ implies that not all of the energy entering the TCSC is conserved (as stored energy), and depending on how the controller parameters are designed, there could be either dissipation or additional production of oscillation energy. The former, improves overall damping and has a stabilizing effect on the system, while the later, can can be a reason for instability. The time-derivative of the second term in $W_{\text{TCSC}}^D(t)$, measures the rate of energy dissipation (or production) in the FACTS device. A negative slope\footnote{the term `slope', in this paper, refers to the average slope over a cycle (i.e, slope of the dc trend)} of $W_{\text{TCSC}}^D$ signifies, the TCSC is behaving as a source of oscillation energy. Similarly, a positive slope$^3$ of $ W_{\text{TCSC}}^D$ signifies, the TCSC is dissipating oscillation energy (i.e., behaving as a sink). Observe that, for the controller structure in Fig  \ref{controller}, the time-derivative of $W_{\text{TCSC}}^D$ (see, \ref{wtcsc_def_2}) is proportional to the controller gain $K_p$. Therefore, an incorrect setting of $K_p$ can transform a TCSC from a sink to a source of oscillation energy. This is verified with the case study presented next.

\begin{figure}[h]
    \centering
    \vspace{-0.2cm}
    \includegraphics[width = 0.95\linewidth]{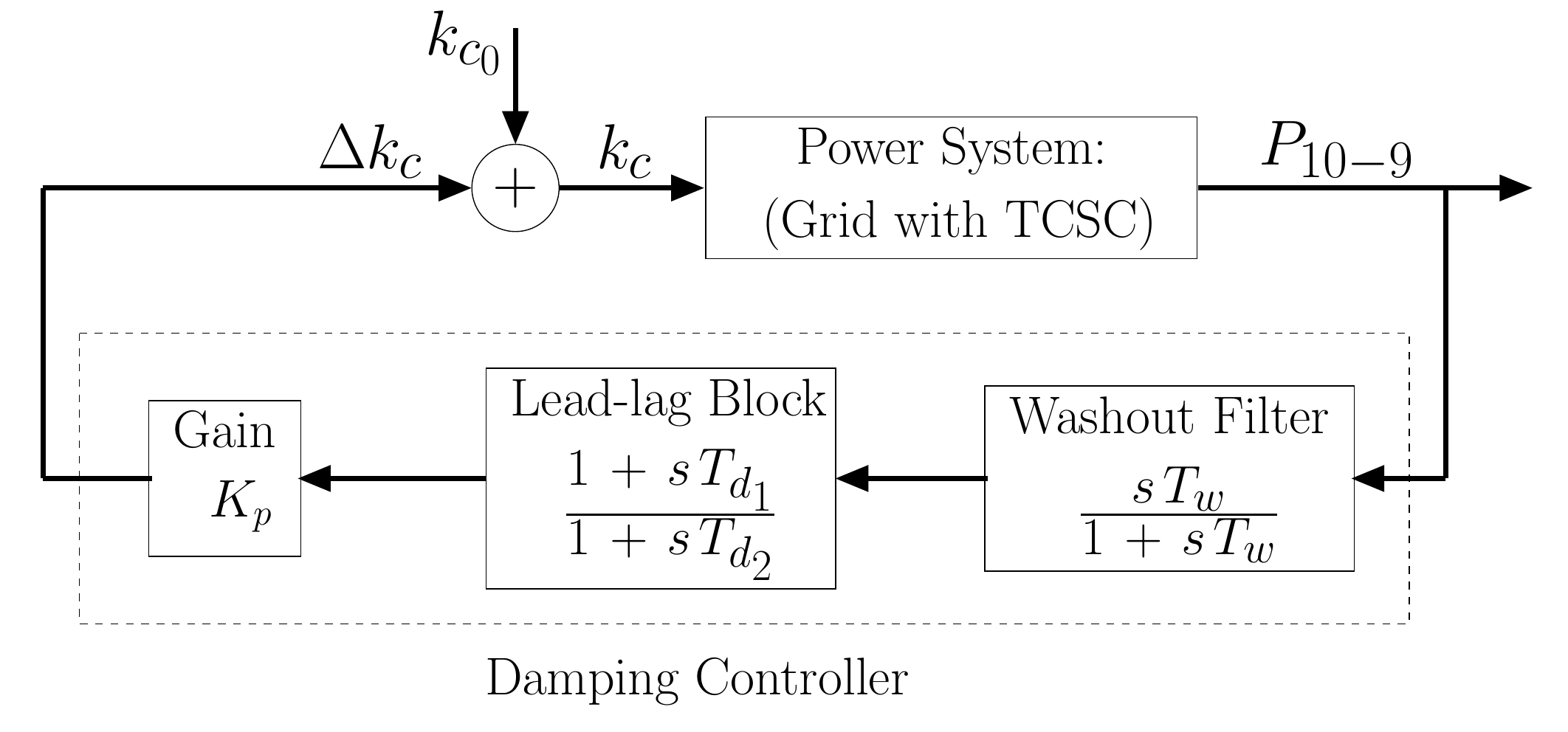}
    \caption{Damping controller modulating $k_c$ of a TCSC.}
    \label{pss}
    \vspace{-0.2cm}
\end{figure}
\begin{figure}[h]
\vspace{-0.4cm}
\begin{subfigure}{\columnwidth}
  \centering
    \includegraphics[width=\linewidth]{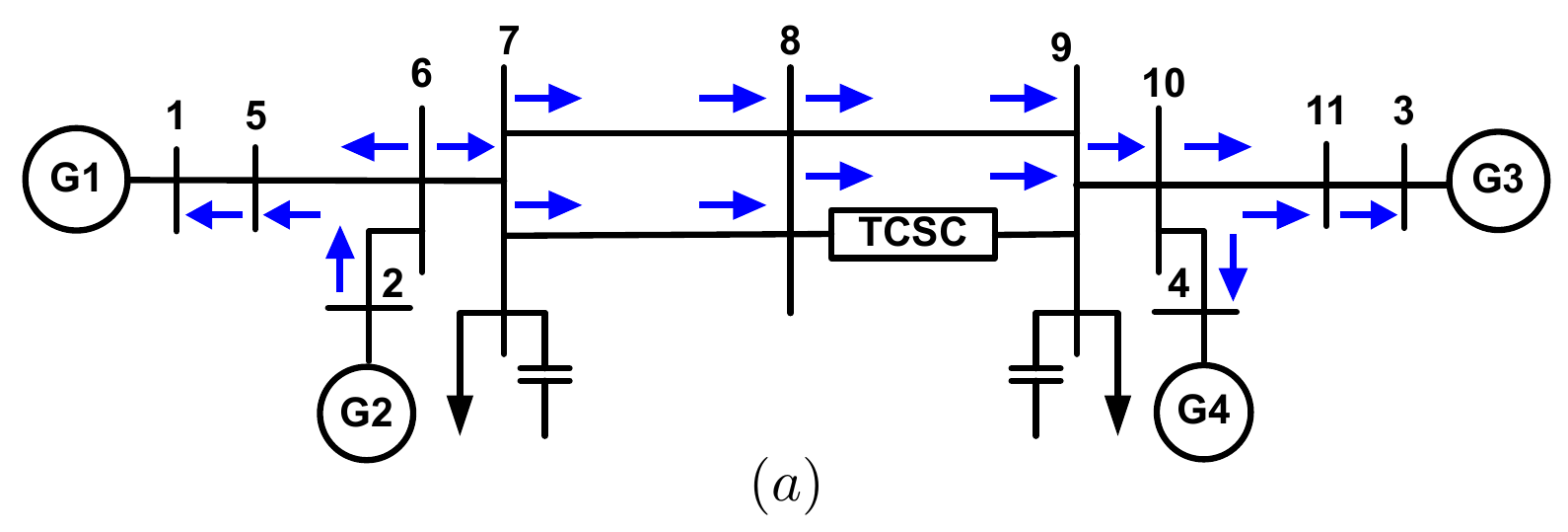}
\end{subfigure}
\vspace{0.1cm}
  \begin{subfigure}{.5\columnwidth}
    \centering
    \includegraphics[width=\linewidth]{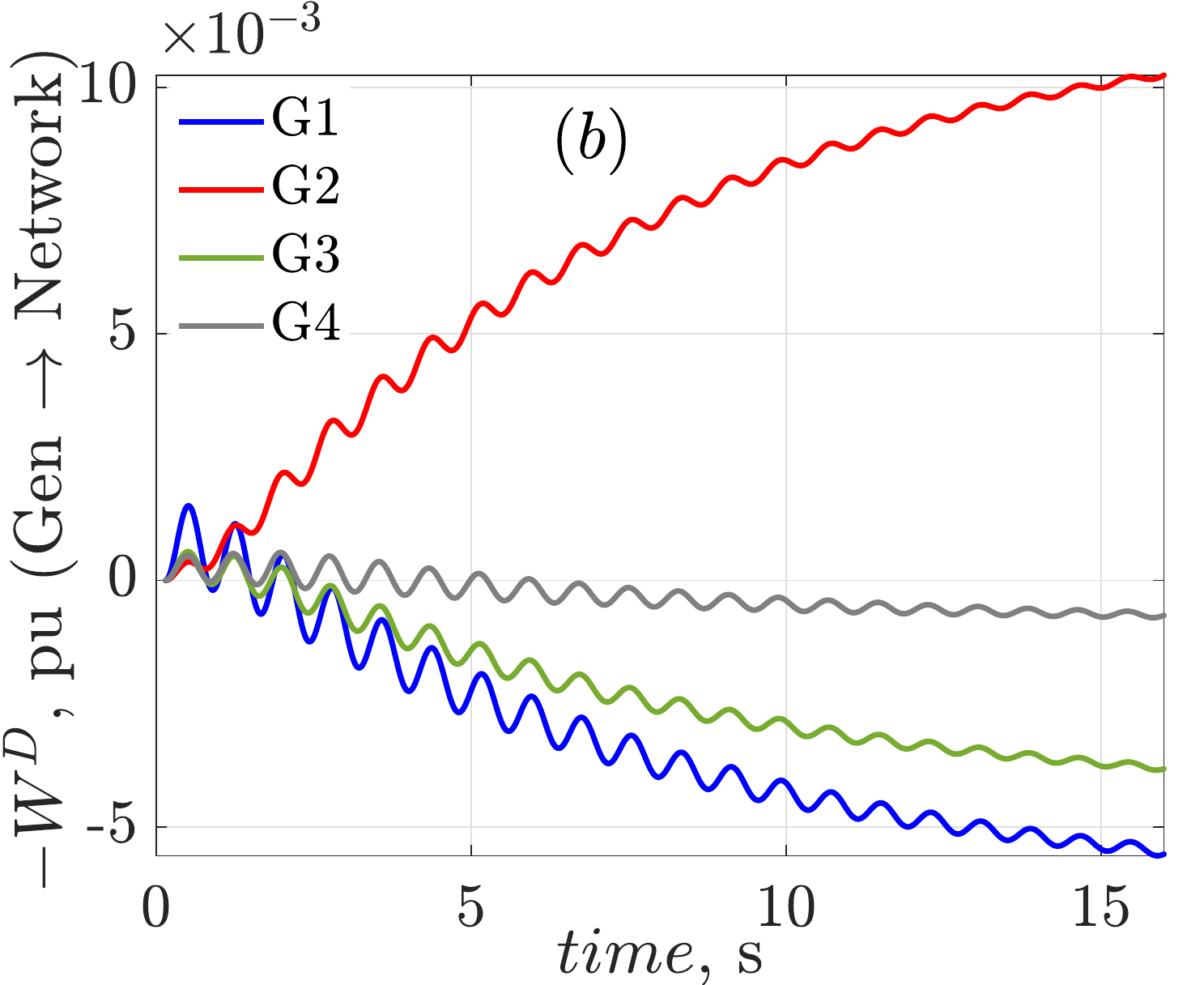}
  \end{subfigure}%
  \begin{subfigure}{.5\columnwidth}
    \centering
    \includegraphics[width=\linewidth]{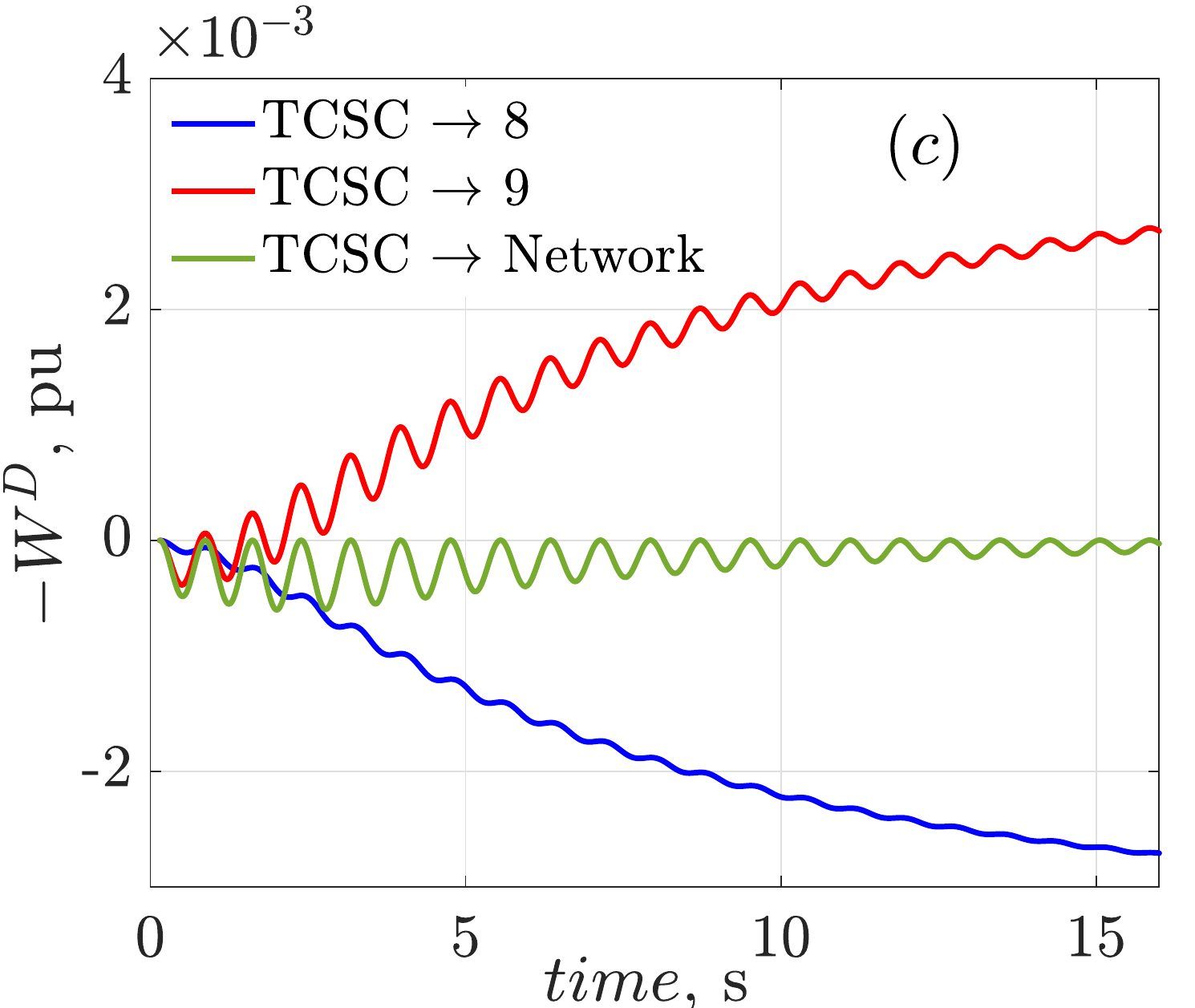}
  \end{subfigure}%
  \caption{$(a)$ Paths of DEF in the network, $(b)$ DEF from generators into the network, and $(c)$ DEF from TCSC into the network, for $K_p = 0$. }
  \label{nominal}
  \vspace{-0.1cm}
  \end{figure}

\textbf{Case Study (A):} Consider the IEEE $2-$area $4-$machine test system \cite{kundur} in Fig \ref{nominal}$(a)$ with the following modifications $-$ (1) a TCSC is introduced in one of the transmission lines connecting buses $8$ and $9$, (2) for the TCSC-connected line, at steady state, $k_{c_0} = 0.3$, and (3) a damping controller (see, Fig  \ref{pss}) is introduced to modulate $k_c$ based on the transients in the feedback signal $P_{10-9}$ (real power flow between buses $10$ and $9$, measured at bus $10$, Fig \ref{nominal}$(a)$). The design parameters of the controller are as follows: $T_{d_1} = 0.4867$ s, $T_{d_2} = 0.0543$ s, and $T_w = 10$ s. Three different DEF scenarios are simulated
by varying  $K_p$ values: $(i)$ $K_p = 0$, implying, $\Delta k_c = 0$ and the TCSC is operating under fixed compensation, $(ii)$ $K_p = 0.0527$, the design value to achieve the desired pole-placement for a well-damped transient response, and $(iii)$ $K_p = -0.0527$, negative of the design value in $(ii)$. 
\par Since, the TCSC is connected to one of the lines between buses $8$ and $9$, we calculate the DEFs out of the TCSC \footnote{by definition (see, (\ref{Wik1}), (\ref{Wik3})), $W^D$ is flowing into the device, and $-W^D$ out of the device}$^{,}$ \footnote{measured at $8$ using the $\Delta P_{ik}$ flowing into $8$ from the TCSC, and at $9$ using the $\Delta P_{ik}$ flowing into $9$ from the TCSC} into the buses $8$ and $9$. The total DEF out of the TCSC into the network is the sum of these two individual DEFs. The results for $(i)$ $Kp =0$ is shown in Fig  \ref{nominal}. Observe from Fig  \ref{nominal}$(c)$,
\begin{equation*}
\begin{aligned}
\vspace{-0.3cm}
  & -\frac{d}{dt}W^D_{\text{TCSC}} ~= ~ \text{slope~of~} \text{DEF}_{\text{~TCSC $\rightarrow$ Network}} \\&=~ \text{slope~of~} \text{DEF}_{\text{~TCSC $\rightarrow$ 8}}~ +~ \text{slope~of~}\text{DEF}_{\text{~TCSC $\rightarrow$ 9}} \approx~ 0.
 \end{aligned}
\end{equation*}
\par \noindent This supports our proposition $1$ that a TCSC with fixed compensation is neither a source nor a sink of oscillation energy. The DEF from all four generators into the network, and the path of DEF in the network are shown in Figs \ref{nominal}$(b)$ and $(a)$, respectively. 

\par
\begin{figure}[h]
\vspace{-0.4cm}
\begin{subfigure}{\columnwidth}
  \centering
    \includegraphics[width=\linewidth]{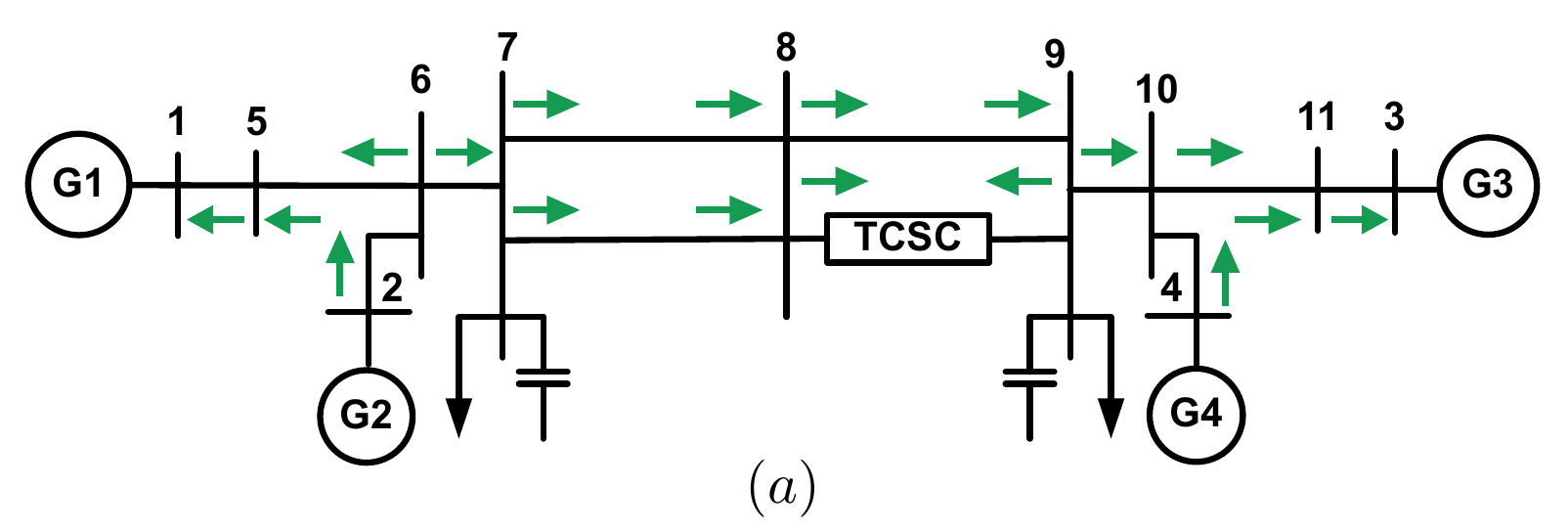}
\end{subfigure}
\vspace{0.1cm}
  \begin{subfigure}{.5\columnwidth}
    \centering
    \includegraphics[width=\linewidth]{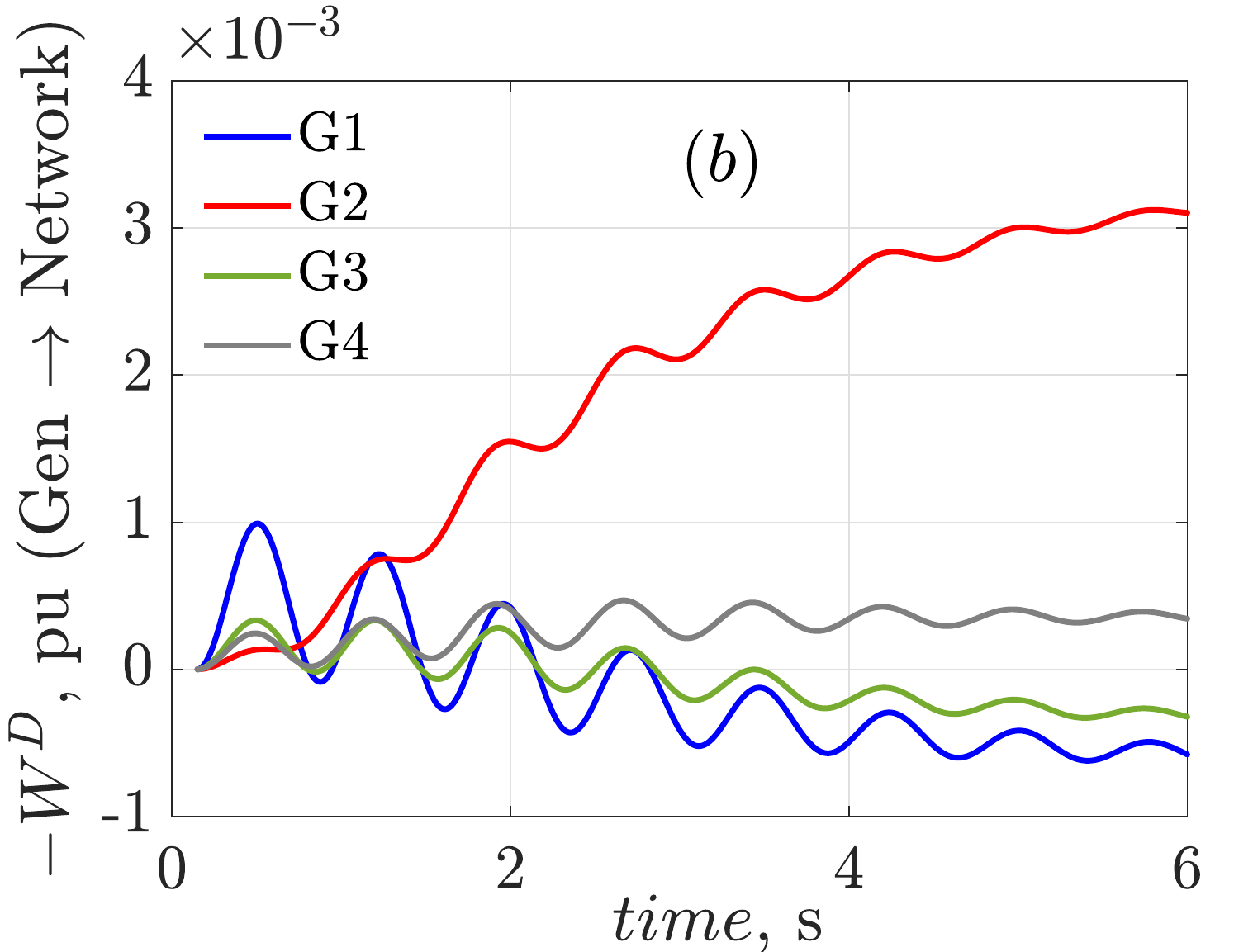}
  \end{subfigure}%
  \begin{subfigure}{.5\columnwidth}
    \centering
    \includegraphics[width=\linewidth]{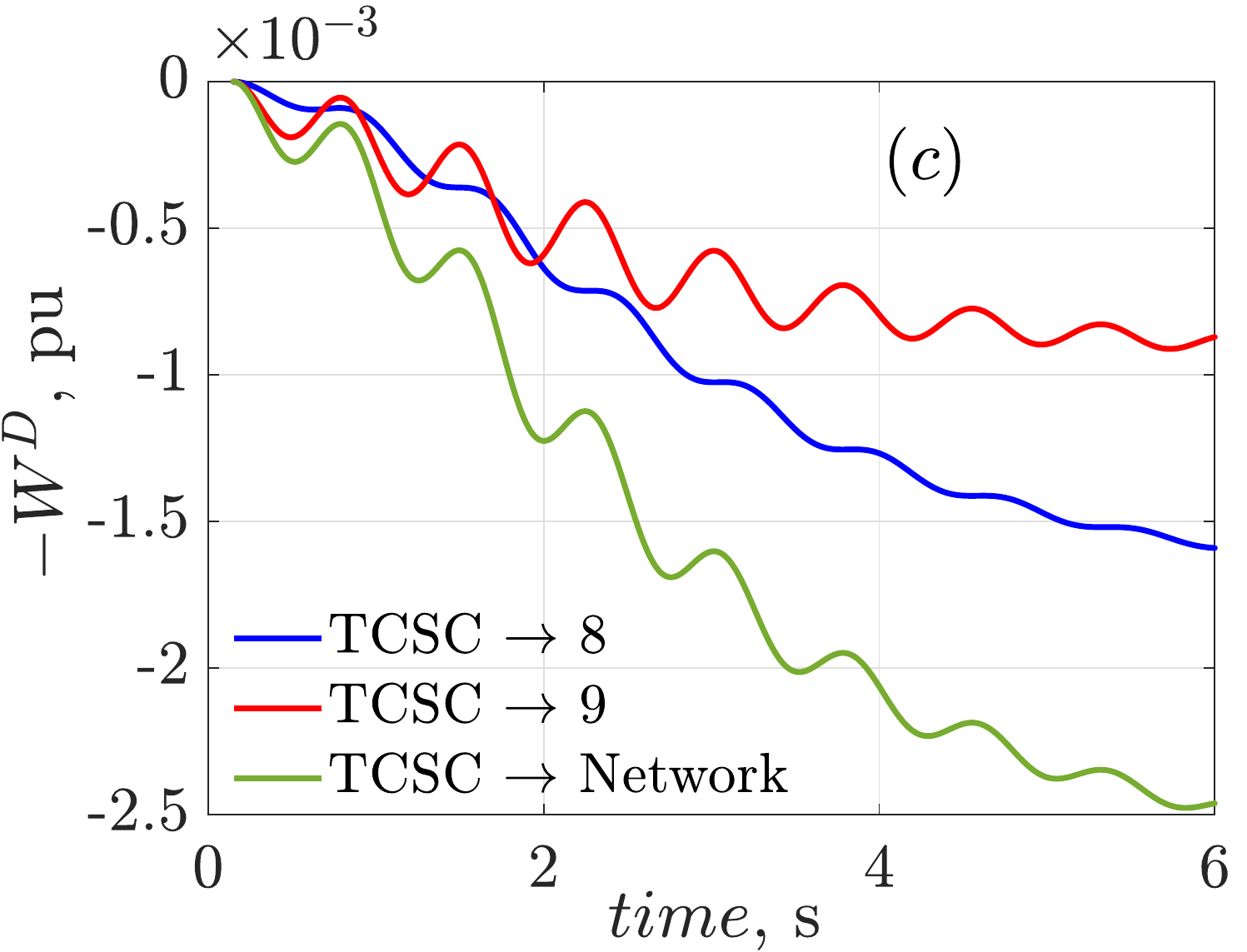}
  \end{subfigure}%
  \caption{$(a)$ Paths of DEF in the network, $(b)$ DEF from generators into the network, and $(c)$ DEF from TCSC into the network, for $K_p = 0.0527$.}
  \label{tcsc_sink}
  \vspace{-0.4cm}
  \end{figure}

\begin{figure}[h]
\begin{subfigure}{\columnwidth}
  \centering
    \includegraphics[width=\linewidth]{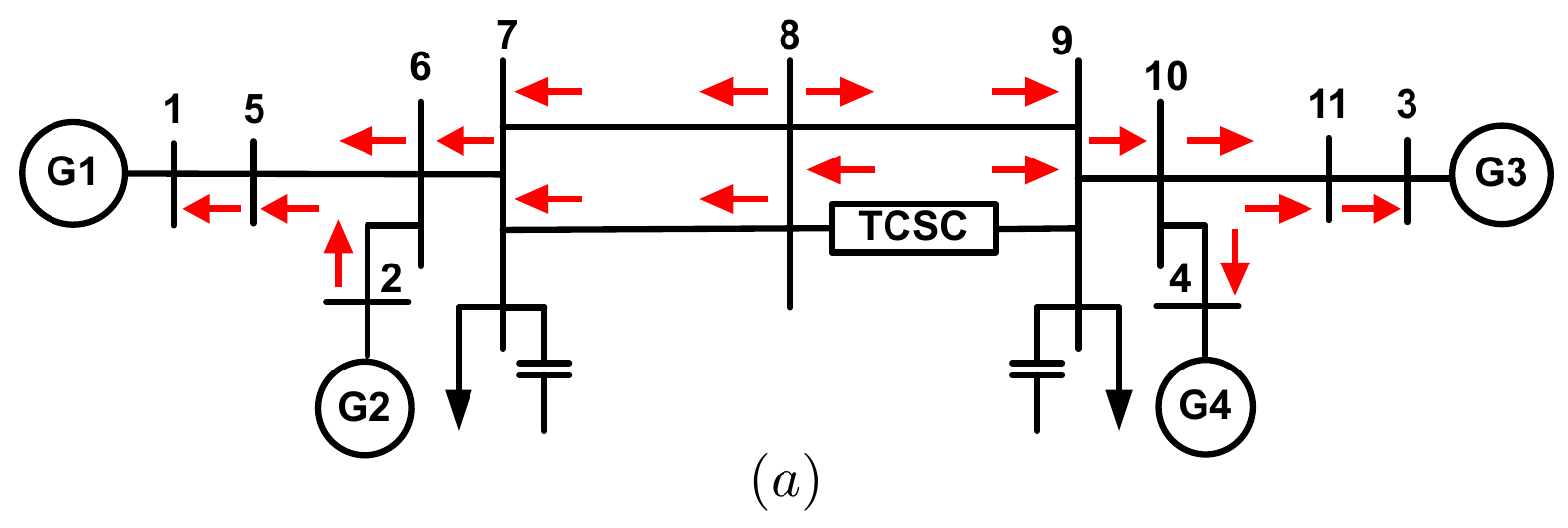}
\end{subfigure}
\vspace{0.1cm}
  \begin{subfigure}{.5\columnwidth}
    \centering
    \includegraphics[width=\linewidth]{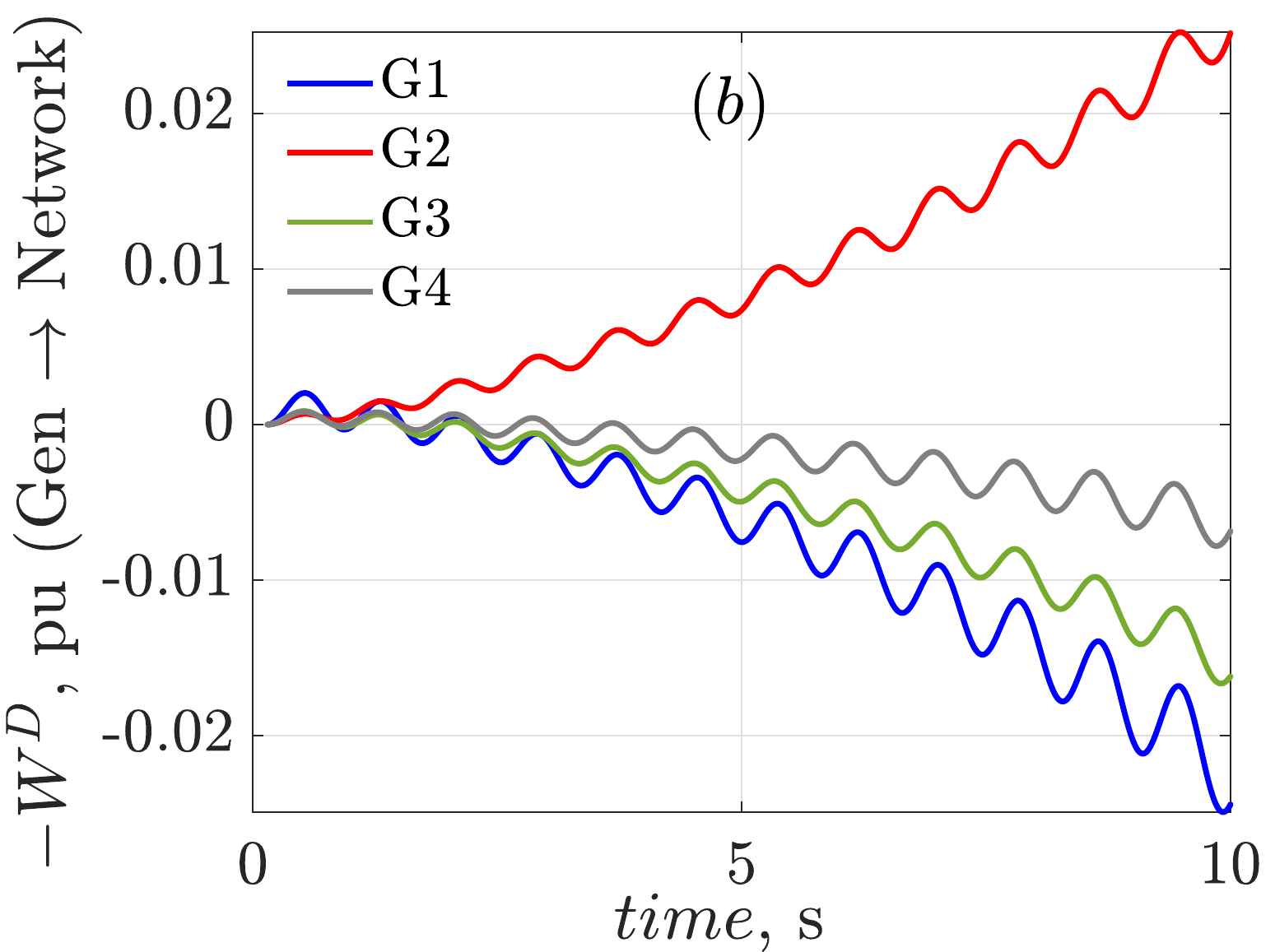}
  \end{subfigure}%
  \begin{subfigure}{.5\columnwidth}
    \centering
    \includegraphics[width=\linewidth]{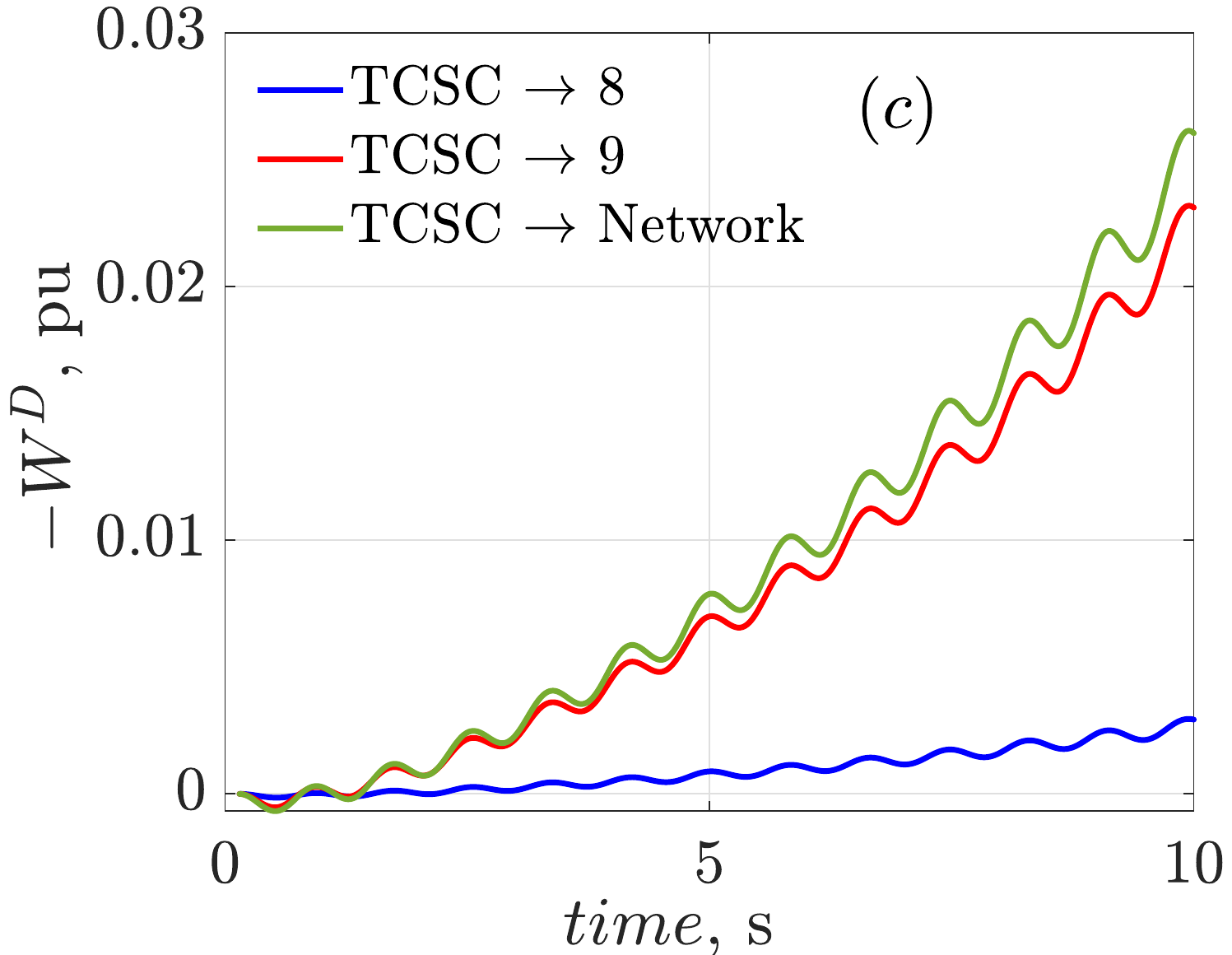}
  \end{subfigure}%
  \caption{$(a)$ Paths of DEF in the network, $(b)$ DEF from generators into the network, and $(c)$ DEF from TCSC into the network, for $K_p = -0.0527$.}
  \label{tcsc_source}
  \vspace{-0.3cm}
  \end{figure}

 Next, we study scenario $(ii)$, where a damping controller modulates $k_c$ to achieve a desired damping.  Observe from Fig  \ref{tcsc_sink}$(c)$ that the slopes of both the DEFs, from TCSC into buses $8$ and $9$ respectively, are negative. Therefore, $\frac{d}{dt} W^D_{\text{TCSC}} > 0$, implying, the TCSC is dissipating oscillation energy. This is intuitive, since, the controller was designed to add damping to the modal oscillation. The DEF from the generators into the network, and the path of DEF in the network are shown in Figs. \ref{tcsc_sink}$(b)$ and $(a)$, respectively.  Finally, in scenario $(iii)$ we reverse the sign of gain $K_p$ of the damping controller. As seen from Fig  \ref{tcsc_source}$(c)$, the slopes of DEF from TCSC to buses $8$ and $9$ are both positive, implying, $\frac{d}{dt} W^D_{\text{TCSC}} < 0$. Therefore, reversing the controller gain alters the nature of the TCSC from sink to source of oscillation energy.

  \vspace{-0.1cm}
  
  \section{DEF Analysis of a STATCOM}
  \label{stat_sec}
 Assume a lossless static synchronous compensator (STATCOM), with an internal node $k$, is connected to the power network at bus $i$.  $\vec{V}_i (= V_{d_i} + jV_{q_i})$ and $\vec{I}_{ik} (= I_{d_{ik}} + jI_{q_{ik}})$ are respectively the phasors of the bus voltage and the current  injected into the STATCOM, in its controller's $d-q$ reference frame. The transient energy flowing into the STATCOM is then given by $ W_{\text{STCM}} = \int \Im( \vec{I}^*_{ik} d\vec{V_i})$. Since, for a lossless STATCOM, $I_{d_{ik_0}} = 0$, the expression reduces to 
 \begin{equation}\small
 \label{statcomW}
 \begin{aligned}
     &W_{\text{STCM}} = \int \Im\Big((I_{d_{ik}}-j I_{q_{ik}}) (dV_{d_i} + j dV_{q_i}) \Big) \\
     &= -\int I_{q_{ik_0}} d \Delta V_{d_i} + \Big(\int \Delta I_{d_{ik}} d \Delta V_{q_i}-\int \Delta I_{q_{ik}} d \Delta V_{d_i}\Big)\\
     &\overset{\Delta}{=} W_{\text{STCM}}^0 + W_{\text{STCM}}^D 
     \end{aligned}
 \end{equation}
where, $W_{\text{STCM}}^0 = -\int I_{q_{ik_0}} d \Delta V_{d_i}$ and $W_{\text{STCM}}^D = \int \Delta I_{d_{ik}} d \Delta V_{q_i}-\int \Delta I_{q_{ik}} d \Delta V_{d_i}$. Clearly, $W_{\text{STCM}}^0$ is path-independent. Further, for constant $I_{ik}$ (implying, $\Delta I_{q_{ik}} =0$ and $\Delta I_{d_{ik}} =0$), $W_{\text{STCM}}^D$ is zero.  Therefore, for this mode of operation, the STATCOM cannot dissipate oscillation energy. This is presented in proposition 4.
 \vspace{0.1cm}
 \par \textbf{Proposition 4.} \emph{A STATCOM operating under constant current control is neither a source nor a sink of oscillation energy.}
 \vspace{0.1cm}
 \noindent This will be verified with numerical studies later in the paper.
 
 \par
\begin{figure}[h]
     \centering
     \vspace{-0.3cm}
     \includegraphics[width=0.95\linewidth]{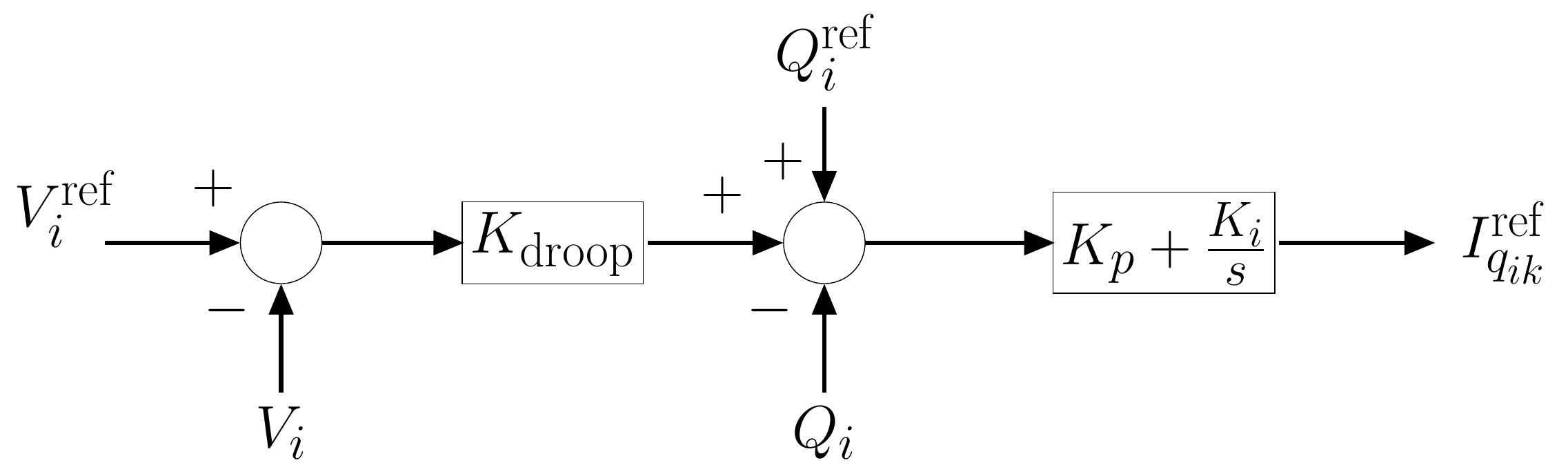}
     \caption{Reactive power control in a STATCOM using $\Delta Q_i^{\text{ref}} - \Delta V_i$ droop.}
     \label{qcontrol}
 \end{figure}

 Next, we study the path-dependence of $W_{\text{STCM}}^D$ for other control strategies. To that end, we consider the reactive power control with $\Delta Q_i^{\text{ref}} - \Delta V_i$ droop, as shown in Fig  \ref{qcontrol}. 
\par \noindent A proportional-integral (PI) controller is used to generate the current reference as follows,
 \begin{equation}\small
 \label{iqref}
 \begin{aligned}
    I_{q_{ik}}^{\text{ref}} = \Big(K_p + \frac{K_i}{s}\Big) \Big( Q_{i}^{\text{ref}} - Q_{i}~ +~K_{\text{droop}}~ (V_i^{\text{ref}} - V_{i})\Big) .\\
\end{aligned}     
 \end{equation}
 In (\ref{iqref}), we substitute, $Q_i =  I_{d_{ik}}V_{q_i} - I_{q_{ik}}V_{d_i}$. Further, considering that the inner current control loop in Fig  \ref{qcontrol} is much faster than the outer voltage and reactive power control loops, while analyzing the relatively slower electromechanical transients, we may assume that the tracking of $I_{q_{ik}}$ is instantaneous. Implying, $I_{q_{ik}} \approx I_{q_{ik}}^{\text{ref}}$. Therefore, we re-write (\ref{iqref}) as
 \begin{equation}\small
 \begin{aligned}
 \label{iq}
      &I_{q_{ik}} = \Big(K_p + \frac{K_i}{s}\Big) \Big((- I_{d_{ik}}V_{q_i} + I_{q_{ik}}V_{d_i}) - K_{\text{droop}}V_{i} + K_{\text{const}} \Big)
      \end{aligned}
 \end{equation}
 where, $K_{\text{const}} = Q_{i}^{\text{ref}} + K_{\text{droop}}V_i^{\text{ref}}$. Next, without loss of generality, consider that the phase-locked-loop in the STATCOM's controller ensures that, in steady state, the $d$-axis aligns with the bus voltage phasor, implying, ${V}_{i_0} = V_{{d_i}_0}$ and $V_{{q_i}_0} = 0$. Therefore, linearization of $V_i^2 = V_{d_i}^2 + V_{q_i}^2 \implies V_{i_0} \Delta V_i = V_{{d_i}_0} \Delta V_{d_i} \implies \Delta V_i = \Delta V_{d_i}.$  Further, considering $I_{d_{ik_0}} = 0$, simplifies the linearization of (\ref{iq}) as follows, 
\begin{equation}\label{iq_del}\small
    \begin{aligned}
  \Delta I_{q_{ik}} = \Big(K_p + \frac{K_i}{s}\Big) \Big((I_{q_{ik_0}} - K_{\text{droop}})\Delta V_{d_i} + V_{d_{i_0}}\Delta I_{q_{ik}}   \Big) . 
    \end{aligned}
\end{equation}
From (\ref{iq_del}), observe that,
\par 1) If $K_i = 0$ (implying, only a proportional controller in Fig  \ref{qcontrol}), then, $\Delta I_{q_{ik}}$ is an algebraic function of $\Delta V_{d_i}$ alone. This implies, the integral $\int \Delta I_{q_{ik}} d \Delta V_{d_i}$ is path-independent.
\par 2) If $K_i \neq 0$, the solution $\Delta I_{q_{ik}}(t)$ is an implicit function of both $t$ and $\Delta V_{d_i}(t)$. Therefore, following the discussions preceding proposition 3, the term $\int \Delta I_{q_{ik}} d \Delta V_{d_i}$ is path-dependent. This implies, $W_{\text{STCM}}^D$ is also path-dependent (see, (\ref{statcomW})). This is summarized in the proposition presented next.
\vspace{0.1cm}
\par \textbf{Proposition 5.} \emph{In a STATCOM with $\Delta Q_i^{\text{ref}} - \Delta V_i$ droop, a proportional-integral controller generating the current reference} $I_q^{\text{ref}}$, \emph{renders $W_{\text{STCM}}^D$ path-dependent.}

Path-dependence of $W_{\text{STCM}}^D$ signifies that, the FACTS device, depending on the design of the controller parameters, can either dissipate oscillation energy (aiding in damping)
 or can behave as a source of oscillation energy. The value of the droop coefficient $K_{\text{droop}}$ can influence this source-sink behavior of the STATCOM. This is presented in the proposition $6$ and in the case study following that. 
\par \textbf{Proposition 6:} \emph{Tuning the droop coefficient} $K_{\text{droop}}$ \emph{transforms the STATCOM from source to sink of oscillation energy.} 
\par We validate propositions $4$ and $6$ through the case study, presented next. 
\par
\begin{figure}[h]
     \centering
     \vspace{-0.4cm}
     \includegraphics[width=0.9\linewidth]{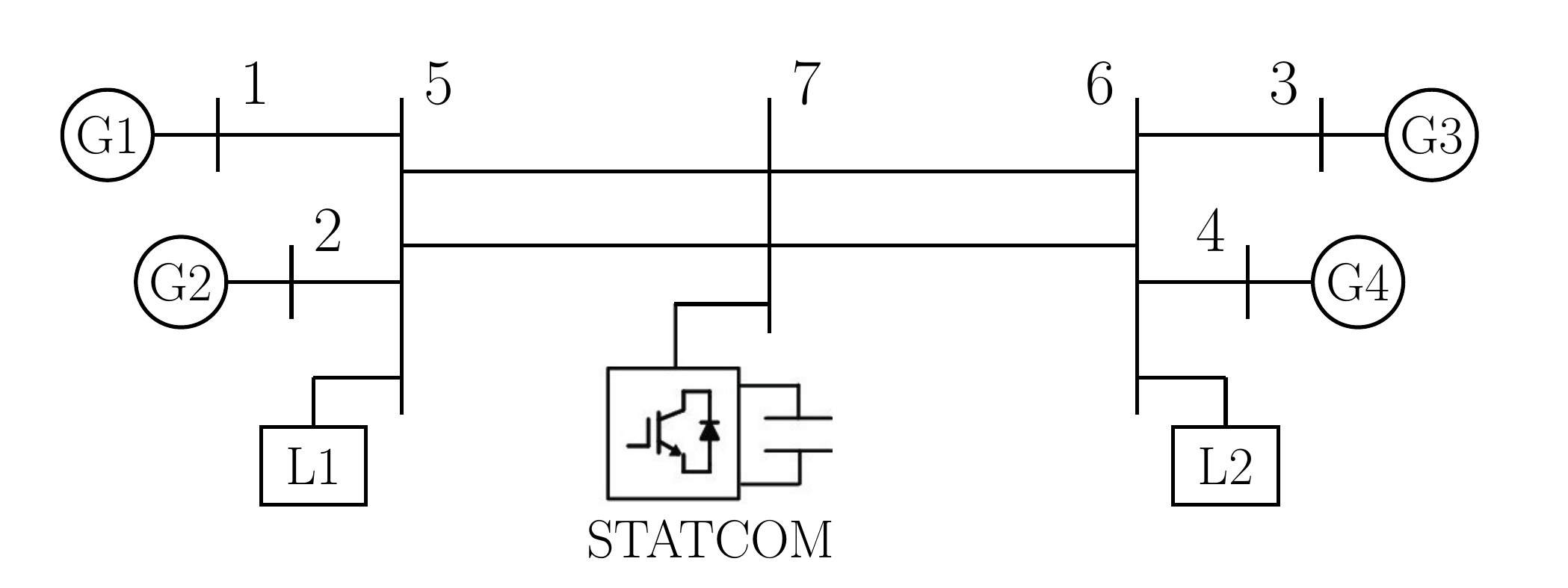}
     \caption{Modified $2$-area $4$-machine system with a STATCOM.}
     \label{4mc_statcom}
 \end{figure}

 \textbf{Case Study (B):} We consider the modified $2$-area $4$-machine system in Fig  \ref{4mc_statcom}. A STATCOM is connected at bus $7$, at the midpoint of the double-circuit tie-line connecting the two areas, as shown. The generator, load, and the line data are same as in \cite{kundur}. 
 \par
\begin{figure}[h]
\begin{subfigure}{.5\columnwidth}
    \centering
    \includegraphics[width=1\linewidth]{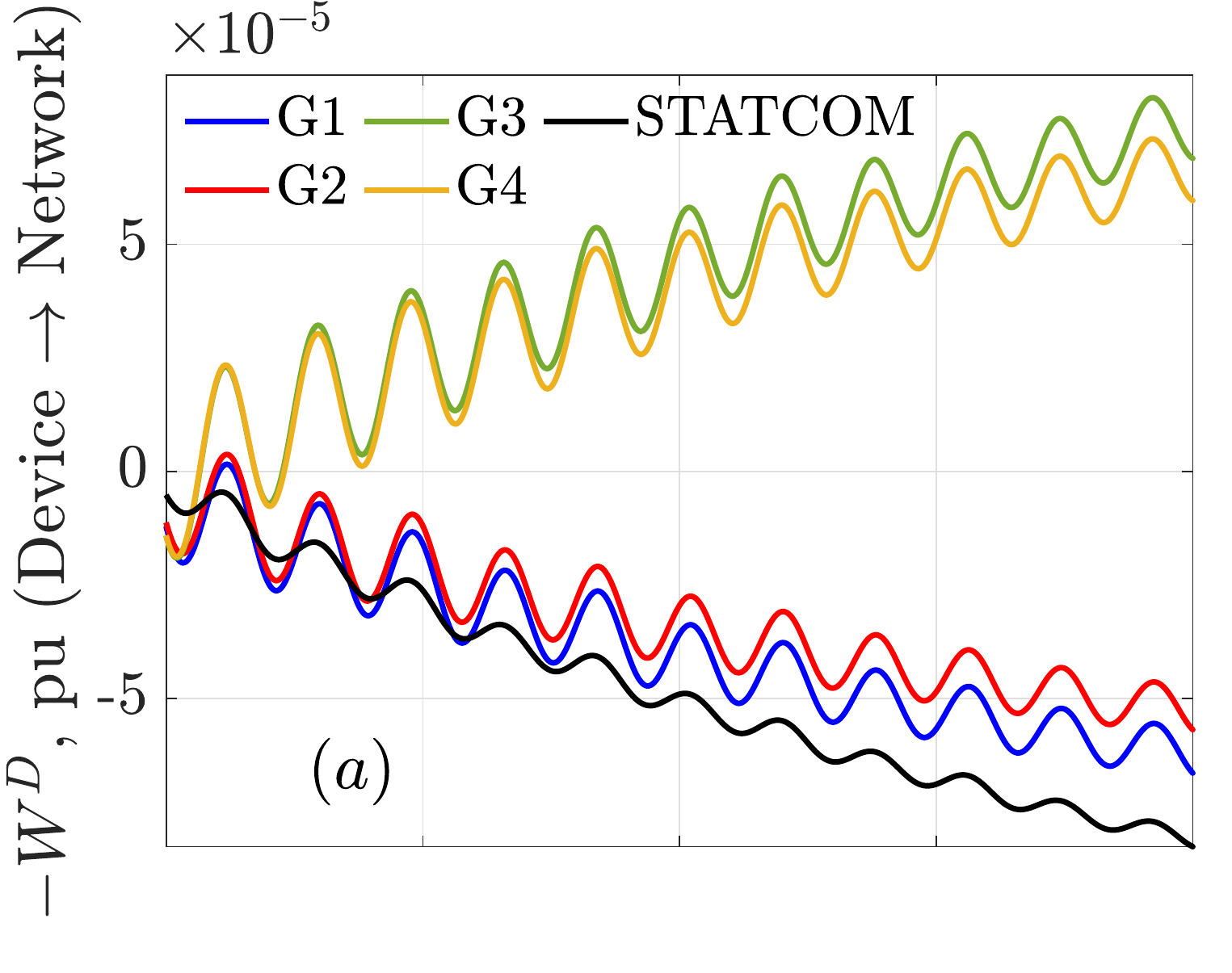}
  \end{subfigure}%
  \begin{subfigure}{.5\columnwidth}
    \centering
    \includegraphics[width=1\linewidth]{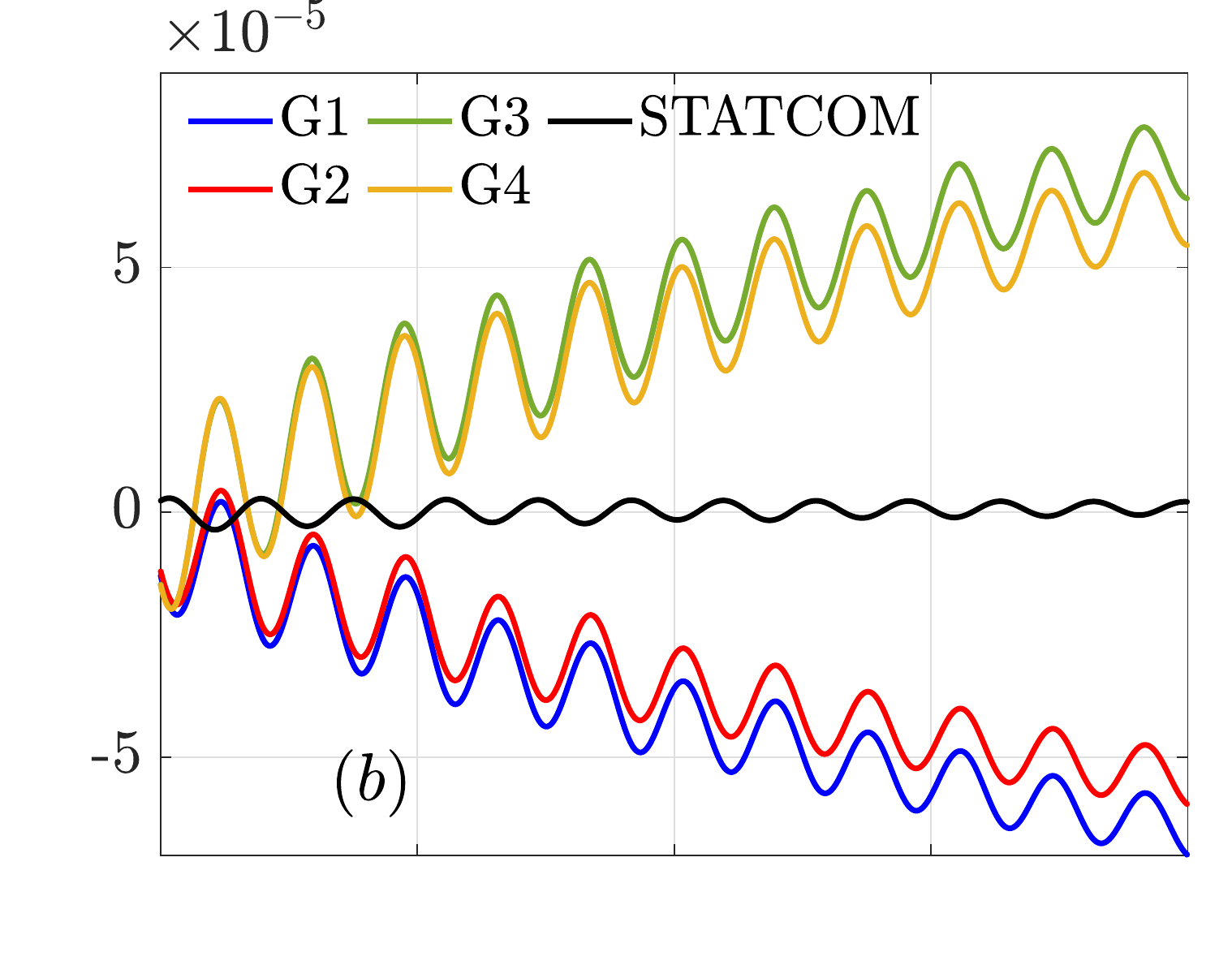}
  \end{subfigure}%
  \\ \vspace{-0.5cm}
  \begin{subfigure}{.5\columnwidth}
    \centering
    \includegraphics[width=1\linewidth]{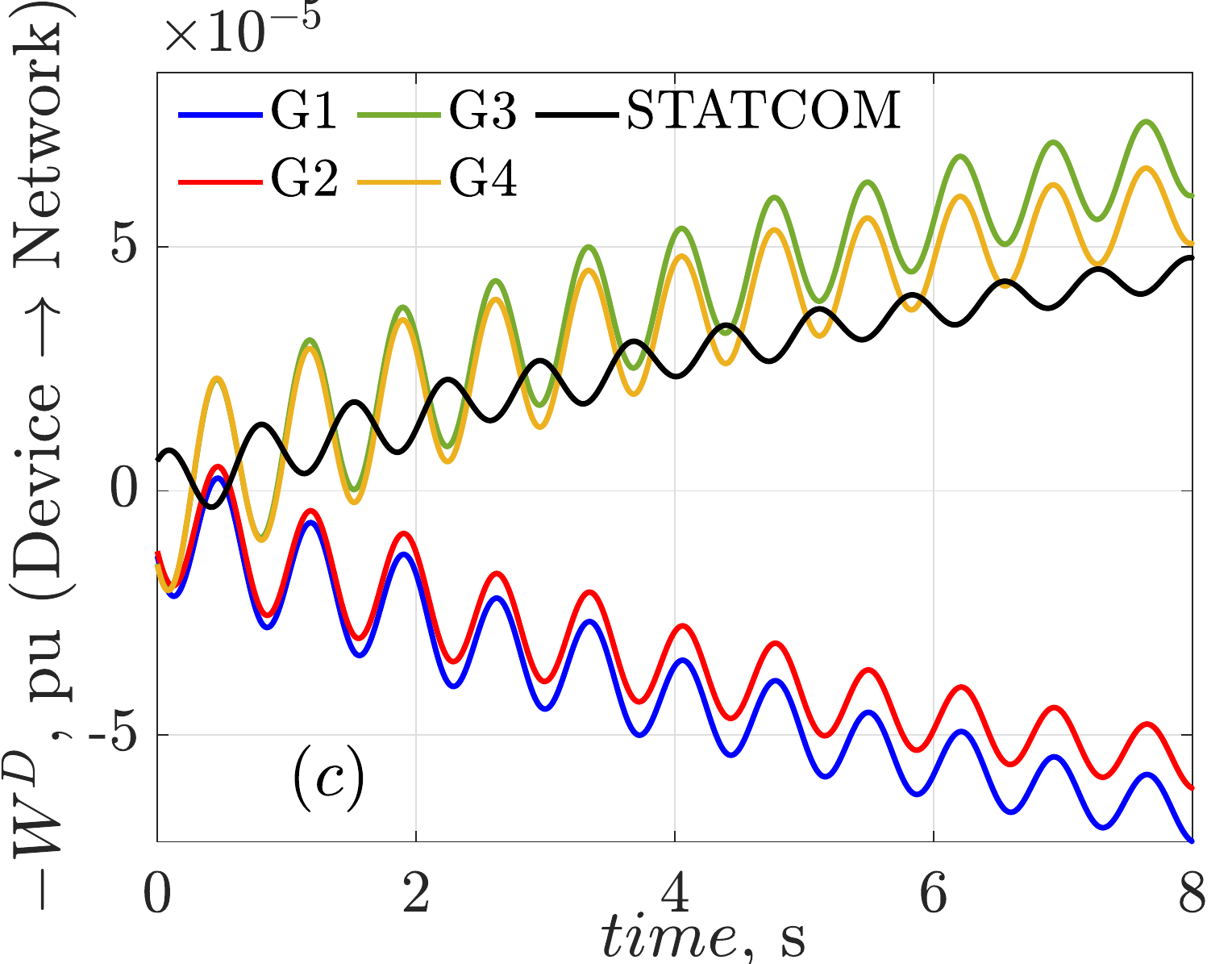}
  \end{subfigure}%
  \begin{subfigure}{.5\columnwidth}
    \centering
    \includegraphics[width=1\linewidth]{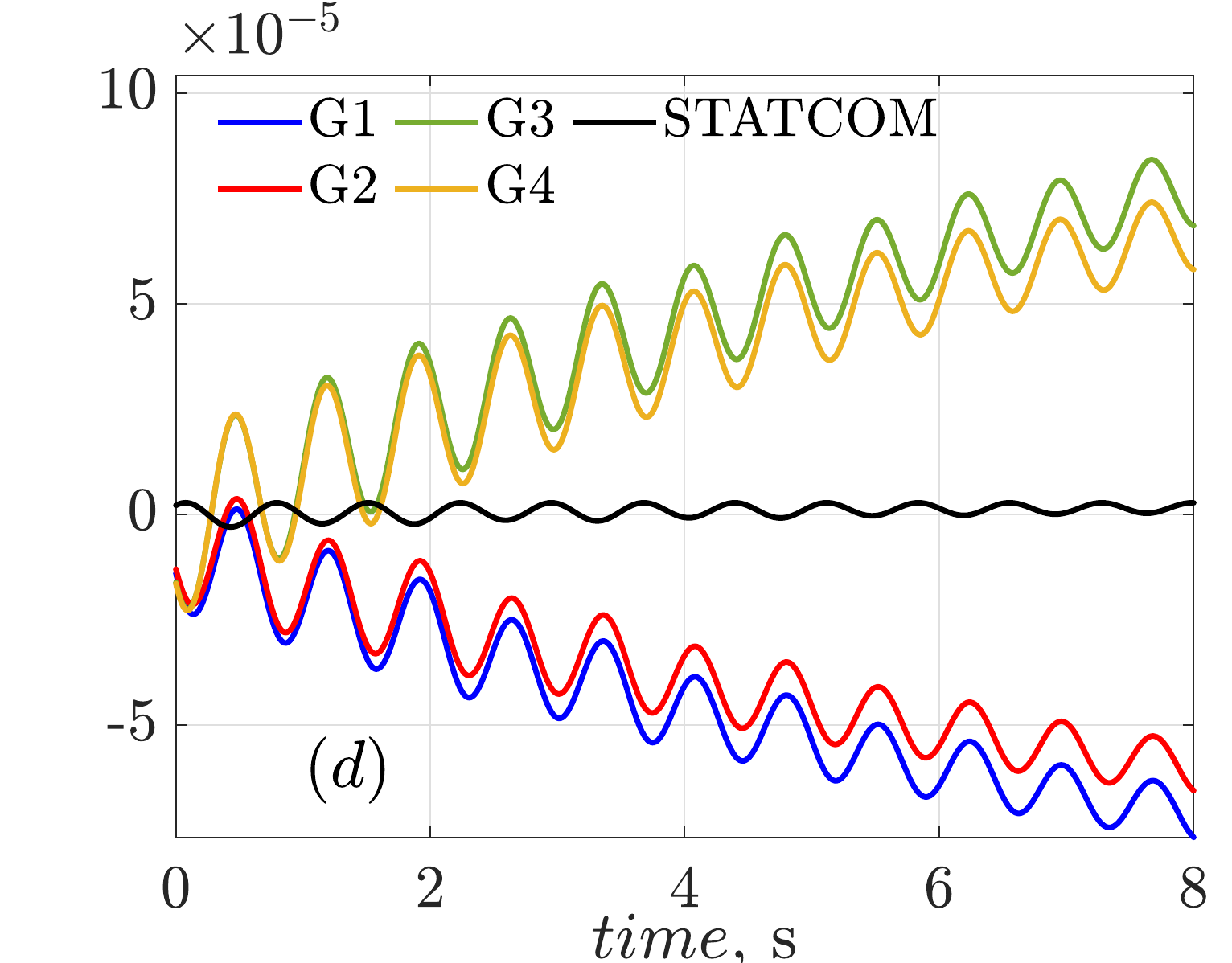}
  \end{subfigure}%
  \vspace{0.6cm}
  \caption{ DEF from the STATCOM and the generators into the network for $(a)$ $K_{\text{droop}} = -1$ VAr/kV, $(b)$ $K_{\text{droop}} = 1$ VAr/kV, $(c)$ $K_{\text{droop}} = 3$ VAr/kV, $(d)$ constant $I_q^{\text{ref}}$.}
  \label{statcom_res}
  \vspace{-0.6cm}
  \end{figure}

 First, we study the DEF for the STATCOM considering the reactive power control strategy in Fig  \ref{qcontrol}. The DEF from the generators and the STATCOM into the network, for different values of $K_{\text{droop}}$, are shown in Figs \ref{statcom_res}$(a)-(c)$. 
  \par As seen from Fig  \ref{statcom_res}, for (a) $K_{\text{droop}} = -1$ VAr/kV, $\frac{d}{dt} W^D_{\text{STCM}} > 0$, implying, the STATCOM is dissipating oscillation energy and contributing towards stabilizing the system mode; for (b) $K_{\text{droop}} = 1$ VAr/kV, $\frac{d}{dt} W^D_{\text{STCM}} = 0, $ implying, the STATCOM is neither a source nor a sink of oscillation energy and therefore, has no damping contribution; and for (c) $K_{\text{droop}} = 3$ VAr/kV, $\frac{d}{dt} W^D_{\text{STCM}} < 0$, implying, the STATCOM is injecting oscillation energy into the network destabilizing the system. These studies support propositions $5$ and $6$, that a controller with frequency dependent phase and gain introduces path-dependent terms in the energy function (observed from nonzero slopes of $W^D_{\text{STCM}}$) and that by tuning $K_{\text{droop}}$ the damping contribution from a STATCOM can be improved. Finally, in Fig  \ref{statcom_res}$(d)$, the DEFs for constant $I_q$ control are shown. Observe, $\frac{d}{dt} W^D_{\text{STCM}} = 0$. This verifies proposition $4$.  
 \section{Conclusions} 
 \label{conclusions}
 The DEF analysis for a TCSC and a STATCOM was presented in the paper. It was observed that a TCSC operating with a fixed compensation can not be a source (or sink) of oscillation energy. However, a frequency-dependent feedback controller modulating its compensation can introduce dissipative effects. Further, an incorrect gain of the controller can transform the TCSC from sink to source of oscillation energy. For the STATCOM, it was shown that under constant current control, the device is neither a sink nor a source. Under droop control, however, the device could behave either as a source or sink, depending on the droop coefficient. 
 \section*{Acknowledgements}
 Financial support from NSF grants under awards CNS 1739206 and ECCS 1656983 is gratefully acknowledged.

\medskip

\renewcommand*{\bibfont}{\footnotesize}
\printbibliography
\end{document}